\documentclass[final,5p,times,twocolumn]{elsarticle}

\usepackage{amssymb}
\usepackage{nidanfloat}

\usepackage{color}
\usepackage{nccmath} 

\journal{Physics Letters B}

\begin{document}

\begin{frontmatter}



\title{Beam energy dependence of charged pion ratio in $^{28}$Si + In reactions}


\author[label1,label2]{M.~Sako\corref{cor1}\fnref{fin1}}
\ead{sakom@riken.jp}

\author[label1,label2]{T.~Murakami}
\author[label2]{Y.~Nakai}
\author[label1]{Y.~Ichikawa}
\author[label3]{K.~Ieki}
\author[label1]{S.~Imajo}
\author[label2]{T.~Isobe}
\author[label3]{M.~Matsushita}
\author[label3]{J.~Murata}
\author[label2]{S.~Nishimura}
\author[label2]{H.~Sakurai}
\author[label1]{R.D.~Sameshima}
\author[label4]{and E.~Takada}

\address[label1]{Department of Physics, Kyoto University, Kyoto 606-8502, Japan}
\address[label2]{RIKEN Nishina Center for Accelerator-Based Scienece, RIKEN, Saitama 351-0198, Japan}
\address[label3]{Department of Physics, Rikkyo University, Tokyo 171-8501, Japan}
\address[label4]{National Institute of Radiological Sciences, Chiba 263-8555, Japan}

\cortext[cor1]{Corresponding author:}
\fntext[fn1]{Present address: RIKEN Nishina Center for Accelerator-Based Scienece, RIKEN, Saitama 351-0198, Japan}

\begin{abstract}
The double differential cross sections for $^{nat}$In($^{28}$Si, $\pi ^{\pm}$) reactions are measured at 400, 600, and 800 MeV/nucleon. 
Both $\pi^+$ and $\pi^-$ are found to be emitted isotropically from a single moving source. 
The $\pi^- / \pi^+$ yield ratio is determined as a function of the charged pion energy between 25 and 100 MeV. The experimental results significantly differ from the prediction of the standard transport model calculation using the code PHITS. This discrepancy suggests that more theoretical works are required to deduce firm information on the nuclear symmetry energy from the  $\pi^- / \pi^+$ yield ratio.
\end{abstract}


\end{frontmatter}


\section{Introduction}
In recent years, both the fields of nuclear physics and astrophysics have investigated the nuclear symmetry energy $E_{sym}(\rho)$, i.e. the isospin dependent term in the nuclear equation of state. 
The density dependence of $E_{sym}(\rho)$ is important for understanding not only the structure of nuclei far from stability \cite{brown, progress} and the reaction mechanism of heavy-ion collision \cite{reaction, daniel, baran} but also many issues in astrophysics \cite{bethe, lattimer}, such as the structure and composition of neutron stars. 
The behavior of $E_{sym}(\rho)$ in the supra-saturation density region $\rho > \rho_0$, where $\rho_0 \cong$ 0.16 fm$^{-3}$ is the saturation density, attracts much interest from the astrophysics after the discovery of a two-solar mass neutron star \cite{2nstar, 2nstar2}.
Unfortunately, its high-density behavior remains very uncertain in spite of much effort from either experimental or theoretical aspects \cite{isospin}. 
To investigate $E_{sym}(\rho)$ in this region using heavy-ion collisions, several kinds of isospin sensitive observable have been proposed, such as the neutron to proton ratio, $\pi^- / \pi^+$, $\Sigma^- / \Sigma^+$, and $K^0 / K^+$ \cite{progress}.
Among these, the $\pi^- / \pi^+$ yield ratio in an intermediate energy  ($\sim$ several hundred MeV) heavy-ion reaction is considered as the most sensitive observable for $E_{sym}(\rho)$.
The current theory for the intermediate energy heavy-ion reactions \cite{progress} suggests that pions keep information of the participant region of the reaction as they are created predominately through a nucleon-nucleon collision process.
According to the $\Delta$ resonance model\cite{STOCK}, the $\pi^- / \pi^+$ ratio is closely related to the neutron to proton ratio in the participant region, while the neutron to proton ratio is strongly affected by the nature of $E_{sym}(\rho)$ through the dynamical isospin fraction.

Recently,  the experimental $\pi^- / \pi^+$ ratio in Au + Au collisions at intermediate energies, which was reported by the FOPI collaboration \cite{FOPI}, was analyzed with various theoretical models based on the transport theory.
However, the extracted results on $E_{sym}(\rho)$ are rather puzzling, i.e. not consistent with each other.
IBUU04 \cite{IBUU04} and IBL \cite{IBL} predict that a super soft  symmetry energy realizes a larger $\pi^- / \pi^+$ ratio, which matches to the naive expectation.
On the other hand, ImIQMD \cite{ImIQMD} and RBUU \cite{RBUU} predict that a stiff symmetry energy realizes a larger $\pi^- / \pi^+$ ratio.
Furthermore the momentum-dependent nuclear mean field in BUU \cite{NMF1} suggests that the integrated yield ratio between $\pi^-$ and $\pi^+$ has no sensitivity, while the differential ratio as a function of the pion kinetic energy keeps its sensitivity to $E_{sym}(\rho)$.
These confusing predictions might arise from the improper treatment of the neutron and proton effective masses in media. 
Indeed, it has been demonstrated that the differential neutron to proton ratio emitted from heavy-ion reactions in relatively low beam energy regions might be sensitive to $E_{sym}(\rho)$ at low kinetic energies of particles, while it might be sensitive to the effective masses at high kinetic energies of particles \cite{wolter}. 
Hong and Danielewiecz have pointed out that the charged pion production in intermediate-energy heavy-ion reactions is also affected not only by the $E_{sym}(\rho)$ but also by the nucleon effective masses, and that the differential pion ratio as a function of beam energy might lead to disentangle these two effects \cite{NMF1}.

To achieve a comprehensive understanding of the pion production mechanism in intermediate-energy heavy-ion collisions and to establish a milestone towards constraining  $E_{sym}(\rho)$ in supra-saturation densities, we have measured the double differential cross sections for the charged pions emitted from $^{28}$Si + $^{nat}$In  reactions over a wide beam energy range using high-intensity stable nuclear beams.
The results are analyzed phenomenologically to elucidate a general trend and are compared with predictions of a standard transport model, Particle and Heavy Ion Transport code System (PHITS) \cite{PHITS1, PHITS2}.
PHITS is widely used for various purposes including radiation shielding design, dosimetry, radiation therapy and space science. 
PHITS has been also successfully applied to account for the double differential cross sections of charged pions for various targets irradiated by 730-MeV protons \cite{PHITS}. 
Our comparison clearly indicates that there are significant discrepancies between our data and the theoretical calculation.
These discrepancies should be resolved before extracting the nuclear symmetry energy information from the experimental data.  

\section{Experimental setup and data analysis}
The experiment was performed at the PH2 beam-line of the Heavy Ion Medical Accelerator in Chiba (HIMAC) in the National Institute of Radiological Science (NIRS). $^{28}$Si beams were accelerated up to 400, 600, and 800 MeV/nucleon with a heavy-ion synchrotron. Typical beam intensities were about 1$\times$$10^7$ particles per spill in a 3.3 sec cycle. A self-supporting natural indium plate (329 mg/cm$^2$ thick) was placed in a small vacuum chamber located at the end of the PH2-line. 

A multiplicity array consisting of 60 plastic scintillators\cite{NN2010} was placed after the target along the beam line, and covered the mid-rapidity region of the polar angle from $27^{\circ}$ to $45^{\circ}$ with respect to the beam direction. 
This array was used to select the centrality \cite{miller}.
To monitor the beam intensity an ion chamber \cite{torikoshi} filled with air at an atmospheric pressure was set 4 m downstream from the target. 

We developed a pion range counter (PRC) to measure the charged pions, $\pi^+$ and $\pi^-$.
The $\pi^+$ events were measured at a low energy range from 10 to 100 MeV, while the $\pi^-$ events were measured from 25 to 100 MeV.
The range counter consisted of a stack of plastic scintillators, similar to the one used by Chiba et al. \cite{chiba}.
In their setup, they successfully selected low energy $\pi^+$ in intermediate-energy heavy-ion reactions in the presence of a large background, but did not simultaneously extract $\pi^-$ due to experimental difficulties \cite{chiba}. 
However, advances in data acquisition system and a new analysis techniques using a reliable simulation have allowed $\pi^-$ to be measured with a range counter.

The PRC consisted of 13 layers, where each layer was coupled to a fast photomultiplier tube (PMT) at the one end through a light guide. 
Here the 13 layers were numbered from i=1 to 13 beginning from the first trigger counter. 
The first two layers, which were each 2 mm thick, were used for triggering the data acquisition. 
Of the remaining 11 layers, two were 5 mm thick, one was10 mm thick, one was 15 mm thick, and seven were 30 mm thick.
To veto charged particles penetrating the PRC, another plastic scintillator (5 mm thick) was placed behind the PRC.

Solid angle and angular acceptance of the PRC were 10.0 msr and $\pm$2.86$^{\circ}$, respectively.  The measurements were performed at $30^{\circ}$, $45^{\circ}$, $60^{\circ}$, $75^{\circ}$, $90^{\circ}$, and $120^{\circ}$ degrees in the laboratory system for all beam energies, except for the beam energy of 400 MeV/nucleon at $75^{\circ}$. 

To identify the $\pi^+$ events, a double pulse signal arising from the $\pi^+\rightarrow \mu^++\nu_\mu$ decay of stopped $\pi^+$ in one of the layers of the PRC was used. 
It should be noted that the decay $\mu^+$ range of $\sim$ 2 mm was short enough for most of $\mu^+$'s to be stopped in the same layer as their parents. 
To ensure separation of the delayed $\mu^+$ signal from the prompt $\pi^+$ one, whose life time was only 26 ns, we utilized fast clipping of the anode signals of the PMTs similar to \cite{chiba}.
The timing information in each double pulse due to $\pi^+$'s and $\mu^+$'s was recorded by a multi hit time-to-digital converter (TDC) with a 250 ps (rms) resolution.

\begin{figure}[h]
\begin{center}
\includegraphics[width=9cm, height=4.5cm]{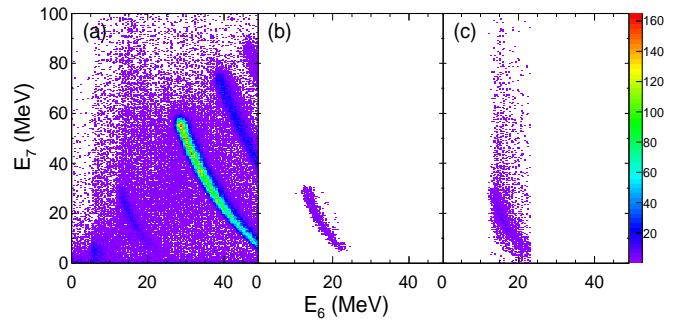}
\end{center}
\caption{(Color online) Correlation of E$_{7}$ vs. E$_{6}$ with a beam energy of 600 MeV/nucleon at 60$^{\circ}$. (a) All events with the condition of S$_7$. (b)$\pi^+$ event with the selection of a double pulse. (c) Charged pion events with the condition of S$_7$ and  G$_7$.}
\label{fig:deltaE}
\end{figure}

On the other hand, $\pi^-$ was more difficult to identify because a stopped $\pi^-$ was absorbed by a hydrogen or carbon nucleus in the plastic scintillator and then disintegrated into many particles in various manners.
This process is referred to as a ''star event''. 
Therefore, the $\pi^-$ yield was estimated in the different way from the $\pi^+$ yield.
In the data analysis, we labeled each triggered event by the last layer that had a hit; when the scintillator counters from the first to the i$^{th}$ layers had hits and the others did not, we assumed that particle stopped in the i$^{th}$ layer, which was labeled the "i$^{th}$ event".
The overall i$^{th}$ event contained both the $\pi^+$ and $\pi^-$ events along with a large background due to other particles like protons, as shown in Fig. \ref{fig:deltaE} (a). 

By requiring double pulse detection for i$^{th}$ events, the $\pi^+$ events were clearly separated from the background for all layers from i = 3$^{rd}$ to 13$^{th}$, except for the trigger layers, as shown in Fig. \ref{fig:deltaE} (b) for i=7.
A fraction of $\pi^+$ events were lost by the selection of the double pulse due to the non-negligible dead time for the double pulse separation. 
The yield of $\pi^+$ stopping in the i$^{th}$ layer was obtained through the integration of the measured time-difference spectra between the $\pi^+$ and $\mu^+$ signals by the well known $\pi^+$ decay half-life of 26 ns. 

The estimation procedure of the $\pi^-$ yield in the i$^{th}$ events is briefly described as follows: 
(1) The yield of all charged pions was evaluated using gate information of the energy deposits in the scintillator. 
(2) The $\pi^-$ yield was obtained by subtracting the $\pi^+$ yield evaluated from the raw yield of the charged pions. 
In this step, two corrections were applied, which are described below. 
(3) The third correction was applied to take the characteristic response of the scintillators due to star events in account.
In the steps (1) and (2), the energy deposit of $\pi^-$ was assumed to be identical to that of $\pi^+$ in the layers where the charged pions passed through.
Because at least four energy-deposit gates were required for clear separation of the charged pions in the present study, we could evaluate the $\pi^-$ yield only from 5$^{th}$ to 13$^{th}$ layers, which corresponded to an energy range of 25 to 100 MeV.

In step (1), the energy deposits ranging from the 1$^{st}$ to (i-1)$^{th}$ layers in the i$^{th}$ events were determined from the $\pi^+$ events.
Then a set of these energy deposit ranges was used as a gate for separate charged pion events from other particles.
Note that we did not use the energy deposit range in the i$^{th}$ layer because $\pi^+$ and $\pi^-$ caused different responses there. 
In addition, the positive correlation of E$_{i-1}$ vs. E$_{i-2}$ was also required to remove the remaining background from other particles, where E$_j$ denoted the energy deposit in the j$^{th}$ layer.
The set of energy deposit ranges and the positive correlation constituted a multi-dimensional energy gate (G$_i$).
Figure \ref{fig:deltaE} (c) shows the result after applying gate G$_7$, where gate G$_7$ successfully reduced the background compared with Fig. \ref{fig:deltaE} (a).

In step (2), two kinds of corrections were taken into account. 
The first one was an excessive selection due to G$_i$, which simultaneously excluded part of the charged pions at the background separation in an acceptable level. 
Thus, the fraction of charged pions reduced by G$_i$ was estimated by comparing the numbers of the $\pi^+$ events with and without G$_i$.
The second one was a correction for mixing charged muons from the in-flight decay of the charged pions.
The events selected with G$_i$ included not only the $\pi^+$ and $\pi^-$ events but also the $\mu^+$ and $\mu^-$ events. 
The muon mixing ratio reached at each layer of the PRC was estimated using the Geant4 simulation.
After these corrections were applied to the yield of the charged pions, they were subtracted to evaluate the $\pi^-$ yield.

In step (3), we estimated the influence of star events in deducing the $\pi^-$ yield.
Particles emitted from star events caused a wide energy deposit distribution, as seen in the E$_7$ direction of Fig. \ref{fig:deltaE} (c). 
Moreover, since some of these events leaked into the neighboring layers and caused additional energy to be deposited in the neighboring layers, the leakage reduced the detection efficiency of the $\pi^-$ events in the i$^{th}$ layer.
The probability of this leak $\alpha_i$ was estimated empirically by considering that the particle emission from a star event was isotropic and by assuming that  the $\alpha_i$ depended on only the layer thickness.
We estimated $\alpha_i$ = 12.1 $\pm$ 0.7 (stat.) $\pm$ 1.9 (sys.)$\%$ for 3-cm thick layer. 
For the other layers, we estimated slightly different values depending upon the thickness. 
These leak probabilities were used to correct the $\pi^-$ yield.

To evaluate the cross section from the yield of the $\pi^+$ and $\pi^-$, the following facts were taken into account:
(1) the correction for the decay in-flight before reaching at the counter ($\sim15 \%$), (2) the correction for losses due to the nuclear absorption or inelastic or elastic scattering by the nuclei in the matter from the target to each layer of the PRC ($1\sim25 \%$ depending on the pion energies), and (3) the correction for the edge effect when pions enter the plastic scintillator with an energy below the threshold (1$\sim$2 MeV depending on the layers) of the discriminator.
Only for a $\pi^+$ event, could a $\mu^+$ particle escape from the scintillator without depositing enough energy to cross over the threshold ($\sim3\%$). 
These correction factors were estimated by Geant4.

Other sources of errors were uncertainties from the beam intensity calibration ($5 \%$) \cite{torikoshi}, the solid angle determination ($1 \%$), and the cross talk estimation ($10 \%$).

\begin{figure*}[b]
  \begin{tabular}{cc}
    \begin{minipage}[b]{0.5\textwidth}
      \includegraphics[width=\textwidth]{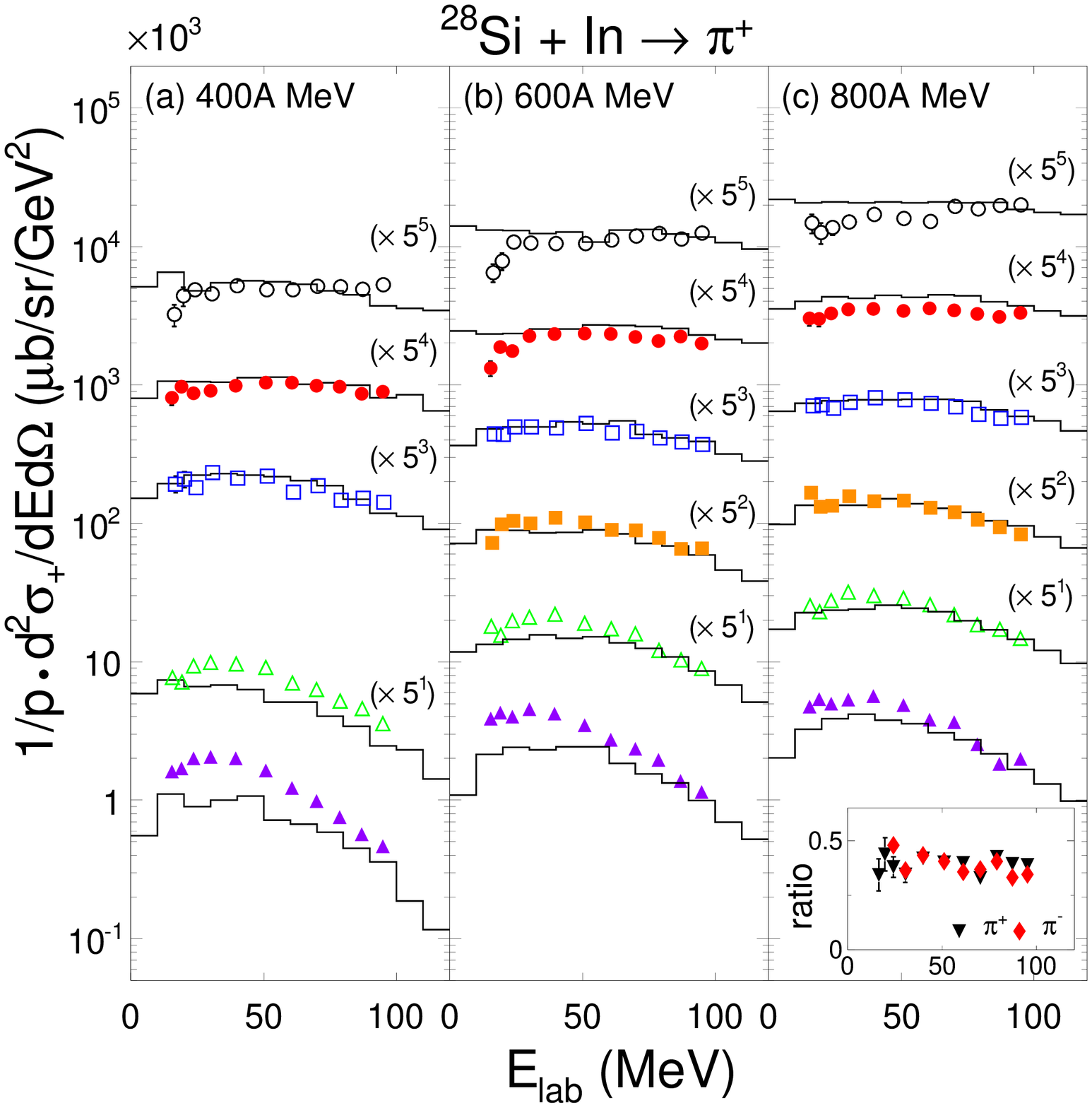}
    \end{minipage}	
    \begin{minipage}[b]{0.5\textwidth}
      \includegraphics[width=\textwidth]{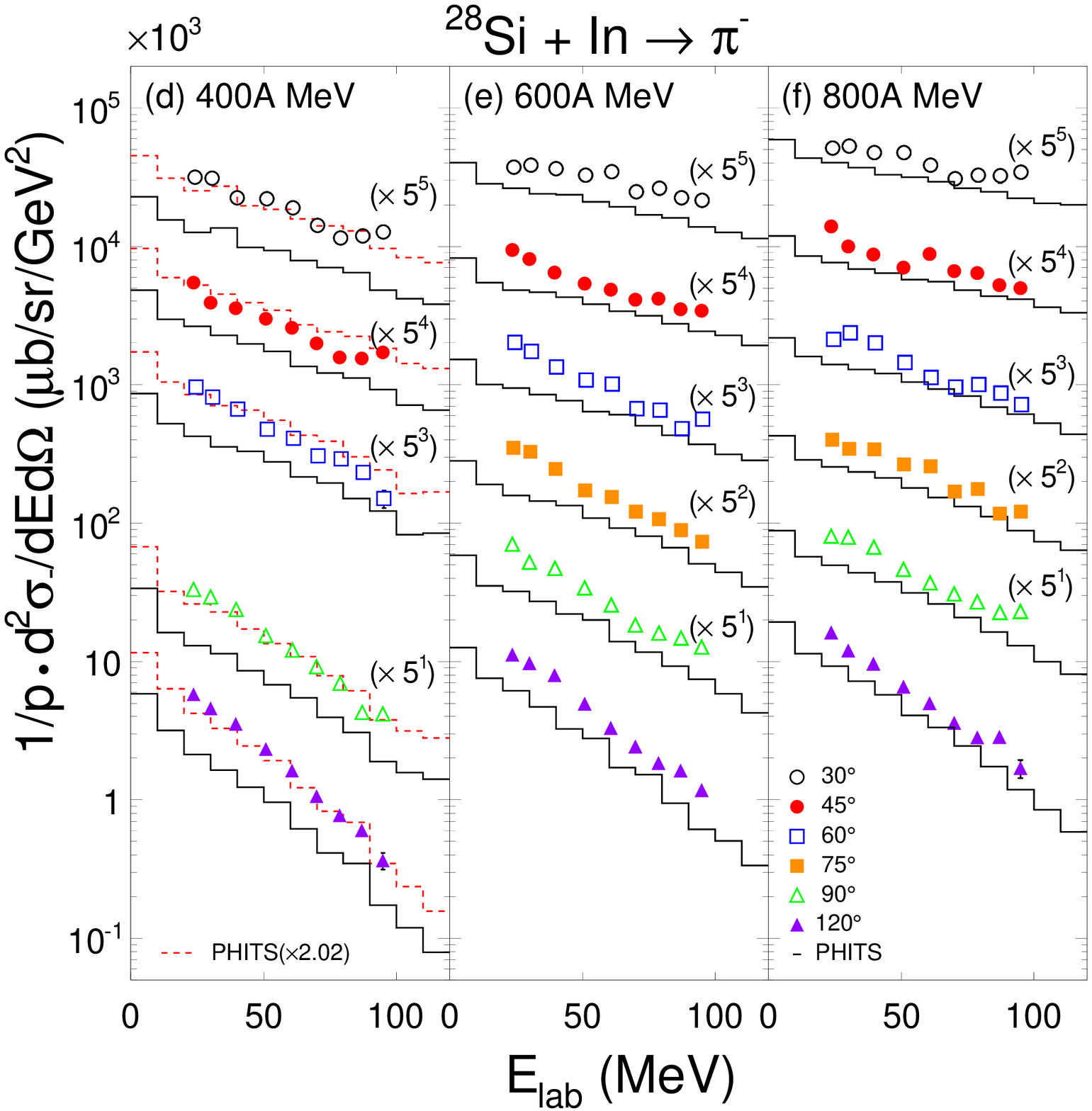}
    \end{minipage}
  \end{tabular}
  \caption{(Color online) Double differential cross sections for the $^{28}$Si + In reactions emitting $\pi^+$ (left panel) and $\pi^-$ (right panel)
with (a, d) 400, (b, e) 600, and (c, f) 800 MeV/nucleon as functions of the charged pion energies ($E_{lab}$).
Markers and solid lines show the experimental data points and PHITS calculation for each angle, respectively.
Dashed histograms of (d) show the PHITS results with a normalized constant.
Inset shows the survival rate of the events using the multiplicity selection with a beam energy of 400 MeV/nucleon at 30$^{\circ}$ for $\pi^+$ (solid circles) and $\pi^-$ (solid square).}
\label{fig:plus}
\end{figure*}

\section{Results and discussion}
Figure \ref{fig:plus} shows the obtained inclusive double differential cross sections of both $\pi^+$ and $\pi^-$ emitted in the $^{28}$Si + In reactions.
The double differential cross sections are for scattering angles of 30$^{\circ}$, 45$^{\circ}$, 60$^{\circ}$, 75$^{\circ}$, 90$^{\circ}$, and 120$^{\circ}$ using the invariant cross section form and as the function of the pion energies at the lab-frame (E$_{lab}$) with beam energies of 400, 600, and  800 MeV/nucleon. 
The energies of the charged pions are estimated from the mean of their maximum and minimum energies stopped in the layers using the Geant4 simulation.

\begin{figure}[h]
\begin{center}
\includegraphics[width=8cm, height=8cm]{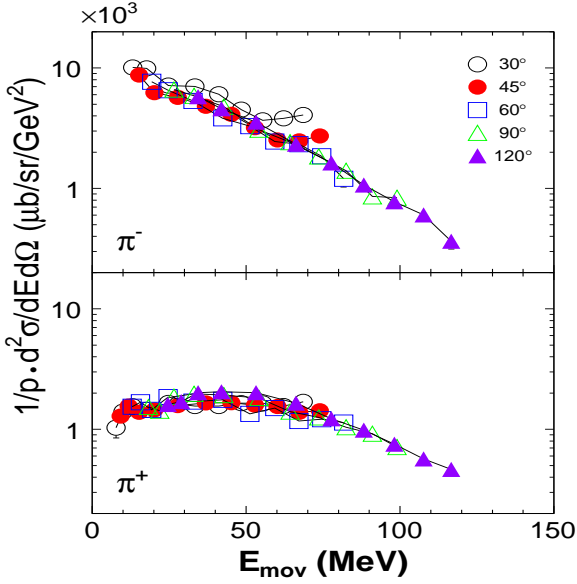}
\end{center}
\caption{(Color online) Invariant double differential cross sections of (top) $\pi^-$ and (bottom) $\pi^+$ for a beam energy of 400 MeV/nucleon in the moving source frame with a 0.43 times mid rapidity.}
\label{fig:moving}
\end{figure}

The cross sections for both charged pions show the expected angular dependences for particles isotropically emitted from a single moving source, namely they decrease exponentially at the high-energy region and their slopes become steeper as the emission angles increase.
In fact, after adjusting the velocities for moving sources, all the invariant cross sections overlap each other rather well.  The resultant velocities are 0.18c, 0.15c, and 0.19c for 400, 600, and 800 MeV/nucleon nucleus-nucleus collisions, respectively. 
These correspond to 43, 31, and 35\% of the mid rapidity and seem to be rather constant. 
A typical example is shown in Fig. \ref{fig:moving} for the 400 MeV/nucleon case.  
Although the energy spectra of $\pi^-$'s  clearly exponentially decrease, the $\pi^+$ energy spectra have a turn over below E$_{lab}$=50 MeV region. 
These resultant velocities of the moving sources are significantly slower than those of the  "usual" mid-rapidity sources and roughly correspond to those of the center-of-mass frame of the entire $^{28}$Si+In system, suggesting that the naive participant-spectator picture might not be applicable to pion production in intermediate-energy heavy-ion reactions.  
It should be noted that although their measurements are limited only to forward angles, Miller et al. found that both $\pi^+$ and $\pi^-$ are isotropically emitted mainly from a single source at rest in the nucleon-nucleon center mass (CM) frame even at sub-threshold energies for pion production from mass symmetric reactions \cite{miller}. 
To resolve such issues in pion production, mass-symmetric systems should also be investigated from the viewpoint of the source velocity in detail because we used rather mass-asymmetric reactions for the current study

The turnover phenomenon in the $\pi^+$ spectra has already been observed by both Chiba et al. and Nakai et al. \cite{chiba, nakai} for the $^{20}$Ne + Pb reactions. 
The angular dependences of their spectra look similar to ours. 
They have claimed that there was a broad bump structure in flow diagrams of the invariant $\pi^+$ cross sections around 90$^{\circ}$ of the center of mass frame for the 800 MeV/nucleon case but not for the 400 MeV/nucleon case, and that such bump structure can be explained only by multiple $\Delta$ formation. 
However, our new data does not hint any existence of such a structure. 
Obviously further investigations on the bump structure shall be performed around beam energies of 400-800 MeV/nucleon.

\begin{figure}[h]
\begin{center}
\includegraphics[width=8cm, height=12cm]{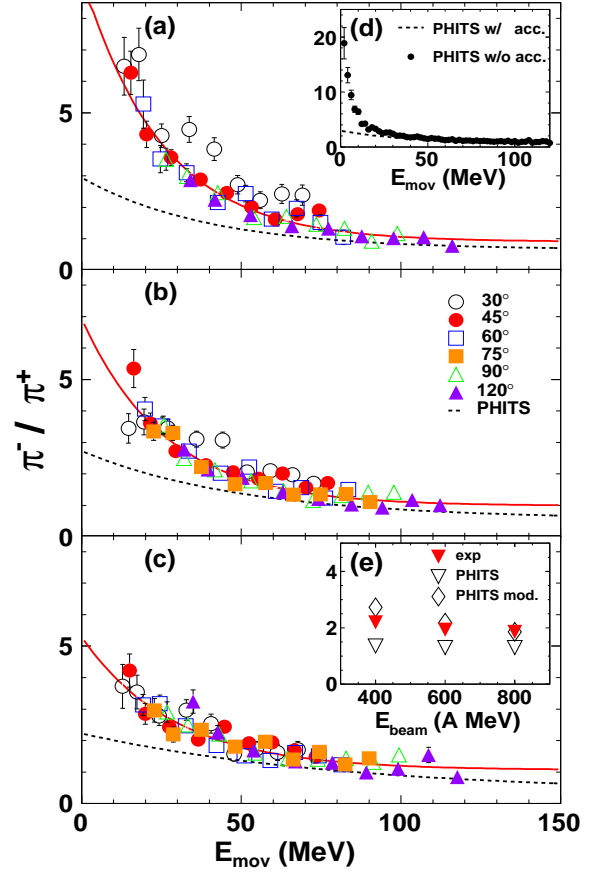}
\end{center}
\caption{(Color online) $\pi^- / \pi^+$ differential yield ratios as functions of the charged pion energies in the moving source frame (E$_{mov}$) with (a) 400, (b) 600, and (c) 800 MeV/nucleon.
Markers show the experimental ratio for each angle and solid (dashed) lines show the fitting results for the experimental (PHITS) ratio in (a) $\sim$ (c).
Differential ratio of the PHITS with (dashed line) and without (solid circle) the experimental acceptance for the beam energy of 400 MeV/nucleon in (d).
Total $\pi^- / \pi^+$ ratio for the experimental data (solid inverted triangles), PHITS within the acceptance (open inverted triangles) and PHITS modified with a normalized constant (open diamonds) in (e) as functions of the beam energy.}
\label{fig:ratio}
\end{figure}

Figure \ref{fig:ratio} shows the $\pi^- / \pi^+$ differential ratio as a function of the energy of the charged pions in a moving source frame (E$_{mov}$) determined for each beam energy of 400, 600, and 800 MeV/nucleon. 
Each data point is the ratio of the inclusive double differential cross sections shown in Fig.\ref{fig:plus}. 
By using the above-mentioned moving source frame, a general trend of the differential ratio of charged pions can be described only by E$_{mov}$, which is independent of the emission angle. 
The ratio approaches a common constant value ($\sim$1) asymptotically at high E$_{mov}$, but increases as E$_{mov}$ decreases. 
The maximum value of the ratio seems to increase as the beam energy decreases. 
The solid curves in the Fig. \ref{fig:ratio} are the results of fitting with a function of $C_1  exp(-C_2  \cdot E_{mov} ) + C_3$, where C$_{1\sim3}$ are the fitting parameters. This functional form can describe the general trend of the $\pi^- / \pi^+$ differential ratio rather well, probably even after integrating over the emission angles. The sets of fitting parameters (C$_1$, C$_2$, and C$_3$) are (8.5, 4.0$\times10^{-2}$, 0.9), (6.0, 3.8$\times10^{-2}$, 1.0), and (4.2, 3.5$\times10^{-2}$, 1.1) for 400, 600, and 800 MeV/nucleon, respectively. Only C$_1$ a shows strong incident energy dependence. 

Usually the central collisions leading to a higher density state show a strong correlation with high multiplicity events. 
Miller et al. have observed, however, that the majority of pion yields for the La + La reaction at 246 MeV/nuceon  \cite{miller} are produced by high-multiplicity (presumably central) collisions, therefore that the selection of pion itself can serve as a good filter for the central events. 
It turned out, in our data, that charged pion events kept their trends in the energy spectra regardless of the associated charged-particle multiplicity measured in mid-rapidity region. 
As a typical example, the ratios between the energy spectra with and without selection of the multiplicity of the upper 50 \% are shown in the inset of Fig.\ref{fig:plus} for the $\pi^+$ and $\pi^-$ with a beam energy of 400 MeV at 30$^{\circ}$. The almost constant ratios indicate only a slight change in the shape of pion energy spectra in terms of charged-multiplicity. 
There is neither significant difference in the angular dependence with and without multiplicity selection. In the present analysis for the charged pions, therefore, we decided not to use information from the multiplicity array to reduce the influence of peripheral events in order to keep sufficient statistics for the pion events. 

In Fig. \ref{fig:plus}, the results of the PHITS calculation (solid histogram) with its default parameters are superimposed as a typical example of a transport model calculation. 
The general trend of the experimental cross sections is roughly reproduced by theory, but there are several striking differences between experimental and theoretical ones. 
In the case of $\pi^+$, the absolute cross sections are rather well reproduced by the theory except for those at backward angles for 400 and 600 MeV/nucleon reactions (underestimation) and those at forward angles for 800 MeV/nucleon reaction (overestimation). 
The PHITS calculation also indicates the turnover structure but the peak positions slightly differ from the experimental ones. 
In the case of $\pi^-$, the slopes of the energy spectra are well explained but the variation of the absolute cross section as a function of the beam energy is much stronger than that of the experimental ones, namely the $\pi^-$ cross section of the 400 MeV/nucleon is significantly underestimated while that of the 800 MeV/nucleon is well reproduced. 
It turned out that one can reproduce the $\pi^-$ cross sections rather well by introducing a single normalization constant to the absolute value of the PHITS calculation for each beam energy. 
The normalization constants are 2.02, 1.68, and 1.43 for 400, 600, and 800 MeV/nucleon, respectively, indicating that the $\pi^-$ yield enhancement depends largely on the incident energy. 
As an example, the dashed histograms in Fig. \ref{fig:plus} (d)  show the results for the 400 MeV/nucleon.
We did not perform a similar attempt for the $\pi^+$ spectra because there is rather significant difference between the experimental and theoretical $\pi^+$ angular distributions.

To understand the origin of the observed differences between the experimental and theoretical angular distributions, we investigated whether the theoretical cross sections can be explained by a single moving source or not. 
For the $\pi^-$ case, we could find the moving frame where all the predicted invariant cross sections overlap each other, but the source velocity is much faster than the experimental one and is close to the "usual" mid-rapidity. 
For the $\pi^+$ case, however, an appropriate moving frame could not be obtained because the absolute cross sections in the turnover region decrease as the angles increase. 
These findings might indicate that the production mechanisms for inclusive $\pi^+$ and $\pi^-$ are quite different in the PHITS calculations.

Although the PHITS results for pions are not consistent with a single moving source picture, we adopted moving sources determined by the analysis of the experimental cross sections, to extract the differential ratio of $\pi^-$ to $\pi^+$ at 30$^{\circ}$, 45$^{\circ}$, 60$^{\circ}$, 75$^{\circ}$, 90$^{\circ}$, and 120$^{\circ}$ from the PHITS calculation. 
Although it might merely be a accidental coincidence, the obtained ratios overlap each other fairly well, except for the 120$^{\circ}$ one.  
A common trend of the theoretical differential ratios is extracted from the overlapped ratios within the experimental acceptance by the fitting procedure used for the  experimental ratios. 
The sets of fitting parameters (C$_1$, C$_2$, and C$_3$) are (2.3, 2.5$\times10^{-2}$, 0.7), (2.2, 2.0$\times10^{-2}$, 0.6), and (1.9, 1.2$\times10^{-2}$, 0.3) for 400, 600, and 800 MeV/nucleon, respectively. 
Unlike the experimental ratio, all the parameters show strong incident energy dependences. 
The results are shown as dashed curves in the Fig. \ref{fig:ratio} (a) $\sim$ (c). 
The theory always underpredicts the ratios, especially at low $E_{mov}$. 

To demonstrate an effect of the acceptance, Fig. \ref{fig:ratio} (d) shows the fitting result for differential ratio predicted by the PHITS calculation integrated within the experimental acceptance together with theoretical differential ratio without having such a constraint for the 400 MeV/nucleon.
The latter ratio shows a huge enhancement at low E$_{mov}$, which is mainly due to the pions emitted towards the backward angles and are out side of the present experimental acceptance. 
This fact clearly indicates that the acceptance should be taken into account carefully when comparing the experimental total ratio to theoretical predictions. 

Because the above comparison, which uses only the empirical moving source even for the theoretical predictions, might be slightly misleading, we also compare the total charged pion ratio within the acceptance of our present experiment. The total ratios are obtained from the double differential cross sections using the following relationship, 

\begin{fleqn} 
\begin{eqnarray*}
\sigma_{-} / \sigma_{+}  &=& \int d \Omega \int dE \frac{d \sigma_{-}}{dE d \Omega}  \Big/ \int d \Omega \int dE \frac{d \sigma_{+}}{dE d \Omega} \\
&\approx& \sum_i sin \theta _i \triangle \theta _i \sum_j \triangle E_j \frac{d\sigma_{-}}{dE d\Omega}  \Big/ \sum_i sin \theta _i \triangle \theta _i \sum_j \triangle E_j \frac{d\sigma_{+}}{dE d\Omega} 
\end{eqnarray*}
\end{fleqn} 
Here, $\sigma_+$ ($\sigma_-$) is the total cross section for $\pi^+$ ($\pi^-$), $\theta _i$ is the measurement angle, and $\triangle$ E$_j$ is the energy acceptance of each layer of the PRC. 

The total ratios of the charged pions obtained from the experimental cross sections and the PHITS calculations within the acceptance are summarized in Fig. \ref{fig:ratio} (e).
The incident energy dependence predicted by PHITS is much weaker than the empirical one. 
This difference in the ratios may arise from the differential cross sections for $\pi^-$. 
In fact, once applying the normalization factors empirically obtained for the $\pi^-$ differential cross sections, the predicted ratios become similar to the experimental ones in terms of the beam energy dependence as shown in the Fig. \ref{fig:ratio} (d) (open diamond).
Taking account of the suggestion by Li et al.\cite{lowpi} in which the multiplicity and spectrum of $\pi^-$ are more sensitive to the symmetry energy than those of $\pi^+$, the normalization factor required to account for the $\pi^-$ cross section could be a good measure for the symmetry energy. 
Since our data seems to require large enhancement factors only for the $\pi^-$ yield, the symmetry energy would be softer than the default setting in PHITS. 
However, since PHITS reproduces the trend of the $\pi^+$ spectra rather poorly, we cannot draw a conclusion that the differences in both the differential and total ratios arise only from the symmetry energy. 
Other effects such as the effective masses of nucleons might affect, at least, the  differential pion ratios \cite{NMF1}. 
It is clear that a solid understanding of pion production in intermediate-energy heavy-ion collisions must be established before constraining the symmetry energy in the supra-saturation density regime. 

In summary, we measured the double differential cross sections of $\pi^+$ and  $\pi^-$ for the $^{28}$Si + In reactions at beam energies of 400, 600 and  800 MeV/nucleons. 
It was found that both $\pi^+$ and  $\pi^-$ are emitted isotropically from the single moving source, whose velocity is quite slower than the mid rapidity.
Additionally, the differential pion ratios represented in such moving frames overlap each other at each incident energy. 
The standard transport model PHITS fails to reproduce the observed absolute cross section, the angular dependence of the cross sections and the charged pion ratio. 
By taking the enhanced normalization factor for the $\pi^-$ yield into account, the incident energy dependence of the total pion ratio  may be reproduced within the experimental acceptance, indicating at least a part of the enhancement might be due to the symmetry energy.
Since the velocity of the moving source cannot be predicted prior to an experiments currently, it is essential to determine the actual emission source frame empirically to obtain the real total pion ratio. 
The theoretical model must be improved to reproduce the features of the differential observables. Experimentally, various observables should be measured to ensure that the extracted information about symmetry energy is accurate.
 
\section*{Acknowledgments}
The authors express their gratitude to Dr. Takeshi Murakami and the staff of the Accelerator Engineering Corporation at HIMAC for their assistance with this experiment.
One of authors (M. Sako) was supported by the Grant-in-Aid for JSPS Fellows (No.24$\cdot$2440), the JRA scholarship from RIKEN and the Grand-Aid for the Global COE Programs "The Next Generation of Physics, Spun from Universality and Emergence".
This work was supported in part by Japanese MEXT Grant-in-Aid for Scientific Research on Innovative Area Granted No. 24105004. 
This work was performed as P226 experiment of Research Projects with Heavy Ions at NIRS-HIMAC.



\section*{References}
\bibliographystyle{elsarticle-num} 
 \bibliography{P226_phys_lett_B_9.10}





\end{document}